\documentclass[12pt]{article}
\usepackage[left=3cm,right=3cm,top=3cm, bottom=2.5cm]{geometry}
\usepackage{graphicx}
\usepackage{enumerate}
\usepackage{color}
\usepackage[super]{natbib}
\usepackage{amsmath}
\usepackage{dsfont}
\usepackage{comment}
\usepackage[markers, nolists]{endfloat}
\usepackage{hyperref}

\newcommand{\given}{\,|\,}

\newcommand{\cov}{{\textrm{cov}}}

\newcommand{\dt}{\textrm{d}t}

\newcommand{\E}{{\mathbf E}} 
\newcommand{\CS}{\pi}
\newcommand{\LM}{\mathrm{LM}}

\newcommand{\HH}{\overline{x}}
\newcommand{\tpred}{s}

\newcommand{\eps}{\varepsilon}

\begin{document}

\title{Landmarking 2.0: Bridging the gap between joint models and landmarking}

\author{Hein Putter, Hans C. van Houwelingen}

\maketitle

\abstract{
The problem of dynamic prediction with time-dependent covariates, given by biomarkers, repeatedly measured over time, has received much attention over the last decades. Two contrasting approaches have become in widespread use. The first is joint modelling, which attempts to jointly model the longitudinal markers and the event time. The second is landmarking, a more pragmatic approach that avoids modelling the marker process. Landmarking has been shown to be less efficient than correctly specified joint models in simulation studies, when data are generated from the joint model. When the mean model is misspecified, however, simulation has shown that joint models may be inferior to landmarking.

The objective of this paper is to develop methods that improve the predictive accuracy of landmarking, while retaining its relative simplicity and robustness. We start by fitting a working longitudinal model for the biomarker, including a temporal correlation structure. Based on that model, we derive a predictable time-dependent process representing the expected value of the biomarker after the landmark time, and we fit a time-dependent Cox model based on the predictable time-dependent covariate. Dynamic predictions based on this approach for new patients can be obtained by first deriving the expected values of the biomarker, given the measured values before the landmark time point, and then calculating the predicted probabilities based on the time-dependent Cox model.

We illustrate the approach in predicting overall survival in liver cirrhosis patients based on prothrombin index.
}

\section{Introduction}
\label{sec:intro}

Biomarkers are commonly used in clinical research and treatment to monitor progression of patients. Prominent examples are PSA in prostate cancer~\citep{pauler2002predicting, taylor2013real}, CD4+ T-cell count and HIV-RNA in HIV-infected individuals~\citep{tsiatis1995modeling, wulfsohn1997joint}, and eGFR in patients with end-stage renal disease \citep{asar2015joint, hu2016joint}. They are used to study the impact of (changes) in the biomarker on disease progression and survival, and to obtain updated prognosis for patients, based on observed marker values, i.e., for dynamic prediction of survival.

Broadly speaking, two approaches are in widespread use for dynamic prediction based on longitudinally measured biomarkers. The first is the use of models that jointly characterize the development of the longitudinal biomarkers and the time to event~\citep{tsiatis2004joint, rizopoulos2012joint}. The advantage of the joint modelling approach is that predictions based on joint models are quite efficient when the model is well specified, and there is software available that can fit these models and produce dynamic predictions in standard situations~\citep{JM, JMbayes}.

The second approach is landmarking~\citep{van2007dynamic, van2011dynamic}, which is a pragmatic approach that avoids specifying a model for the longitudinal markers. The advantage of landmarking is that it is easy to implement; no specialised software is needed to obtain dynamic predictions from landmarking. The disadvantage is that it is less efficient than joint modelling, and can yield small bias when last observation carried forward is used and the biomarkers are coarsely observed.

The objective of this paper is to bridge the gap between joint modelling and landmarking, and develop a method that improves on standard landmarking while avoiding complex integration over random effects, which makes joint modelling computationally demanding.

\section{Notation and common approaches}

We assume that we follow patients, indexed by $i=1,\ldots,n$, from time $t=0$ until an event (called death here) occurs. Let $T_i$ be the time of death, and $C_i$ an independent censoring time; define $\tilde{T}_i = \min(T_i, C_i)$ and the status indicator $D_i = I(T_i \leq C_i)$. There is a continuous biomarker process $X_i(t)$, defined as long as individual $i$ is alive. This process is observed at observation times $t_{ij}$, $i = 1,\ldots,n$, $j=1, \ldots, n_i$. The observation times may be irregular, the number of observations may differ across subjects, but it is assumed that they are uninformative, i.e.~that they do not depend on unobserved marker values or unobserved characteristics. The observations have measurement error or day-to-day variation (white noise), and the actual observed measurements of $X_i(t)$ at $t_{ij}$ are denoted by $x_{ij}$. Other covariates might be present, but will be ignored for the sake of simplicity; they can be included into each of the models we describe below in a straightforward way. The observations are $(\tilde{t}_i, d_i, \mathbf{x}_i)$, with $\mathbf{x}_i = (x_{i1}, \ldots, x_{i,n_i})^\top$.

The objective is to use part of the information of the biomarkers of the patient to estimate the conditional probability that the patient is still alive after a pre-defined time window. More specifically, at a prediction time point $s$ we want to estimate the conditional probability that the patient is still alive at time $s+w$, conditionally on being alive at time $s$ and conditional on the history of the biomarkers up to time $s$, i.e.,
\[
    \CS_i(s+w \given s) = P(T_i > s+w \given T_i \geq s, \HH_i(s)),
\]
with $\HH_i(s)$ denoting the history of all biomarker measurements up to $s$.

A Cox model with a time-dependent covariate $X(t)$
\[
    \lambda(t \given X(t)) = \lambda_0(t) \exp(\beta X(t))
\]
is helpful in understanding biology, but useless in predicting the future. The reason for this is that to obtain $\CS_i(s+w \given s)$, based on information at the prediction time $s$ one would need the future values of $X(t)$ after time $s$. Unless the time-dependent covariate is exogenous one would not know these future values.

\subsection*{Joint modelling}

One way to be able to derive dynamic predictions is to make a model for how $X(t)$ might change over time, given knowledge at the prediction time. For this it is assumed that $X_i(t)$ follows a Gaussian process with mean $\mu_i(t)$, possibly depending on covariates, and covariance function $C(t_1, t_2) = \cov(X_i(t_1), X_i(t_2))$. A popular choice is a linear mixed model like $X_i(t) = \beta_0 + b_{i0} + (\beta_1 + b_{i1}) t$ with fixed effects $\beta_0$ and $\beta_1$ (possibly depending on covariates) and random effects $(b_{i0}, b_{i1})$ assumed to be bivariate normal with mean zero. $X_i(t)$ is observed at $t_{ij}$ with independent measurement errors $e_{ij}$. The standard joint model assumes that the hazard of dying at time $t$ depends on the current value of the biomarker, for instance given by the proportional hazards model
\[
    \lambda(t \given \HH_i(t)) = \lambda_0(t)\exp(\beta \mu_i(t)).
\]
Other options, where the hazard depends on the random effects directly, or on the slope or the area under the curve are also possible. Rizopoulos~\cite{rizopoulos2011dynamic} discusses how to obtain dynamic prediction from such joint models, and software is available in the {\bf{JM}} and {\bf{JMbayes}} packages~\citep{JM, JMbayes}.

Simulation studies have shown that the joint model efficiently estimates the underlying parameters, when the model is correctly specified~\citep{rizopoulos2017dynamic}, and that it is reasonably robust against modest misspecification of the dependence function and against modest deviations of proportional hazards, but that it is quite sensitive to misspecification of the longitudinal trajectory~\citep{ferrer2019individual}.

\subsection*{Landmarking}

Landmarking~\citep{van2007dynamic, van2011dynamic} avoids modelling the marker process. The idea behind landmarking is to select, for a given landmark time point $t_{\LM}=s$, all subjects alive and under follow-up at time $s$. The time-dependent information until time $s$ is summarized in some way. Possibilities to summarize the history are the last observed measurement (last observation carried forward, LOCF), or the last observed measurement and the slope. An extension is to use the ``age'' of the last observation (difference between $s$ and last observed time before $s$) as additional covariate. This summary of the time-dependent covariate is subsequently used in a Cox model in the landmark data set. When interest is in estimating the dynamic prediction probability at $s+w$, it is common to apply administrative censoring at $s+w$. This administrative censoring, or ``stopped Cox''~\citep{van2015comparison} is introduced to make the procedure robust against violations of proportional hazards, although for long term prediction (large $w$) time-varying effects might lead to some bias. A concern is that the staleness (``aging'') of the predictor when using LOCF leads to a mismatch between the true underlying value of the biomarker and the last observation. This measurement error leads to violation of proportional hazards~\citep{van2011dynamic}, which would call for modelling time-varying effects, but it is challenging to find adequate models while at the same time avoiding the threat of overfitting.

A number of approaches have been proposed to improve the simple LOCF landmarking approach. One of them is a two-stage approach~\citep{sweeting2017use, paige2017use, paige2018landmark}, where the data of the time-dependent covariate(s) before the landmark prediction time-point $s$ are used and a mixed model is fit to those data (or all data). Then the Empirical Bayes best linear unbiased predictor (BLUP) is used as a predictor at $s$. It is called ``error free'' but that could be too optimistic. It partly solves the staleness problem of the predictor at $t_{LM}$, but does not take care of the problem that the effect of the last measured biomarker before time $s$ typically becomes smaller as $t$ gets more removed from $s$, leading to a decay of the $\beta(t)$.

\section{Landmarking 2.0: getting closer to the joint model}

The joint model approach leads to the following model for the conditional survival~\citep{putter2017understanding}:
\begin{equation}
\label{eq:central}
    \CS_i(s+w \given s) =   \E \Biggl[ \exp\Bigl(-\int_s ^{s+w} \lambda_0(t)\exp(\beta X_i(t)) \dt \Bigr) \given T_i \geq s, \HH_i(s) \Biggr].
\end{equation}
Following Tsiatis et al.~\cite{tsiatis1995modeling} in their treatment of measurement errors in survival analysis, see also~\cite{andersen2003attenuation}, the conditional survival can by approximated by
\begin{equation}
\label{eq:centralappr}
    \CS_i(s+w \given s) \approx \exp\Bigl(-\int_s ^{s+w} \tilde{\lambda}_0(t)\exp\bigl(\tilde{\beta} \E[X_i(t) \given T_i \geq t, \HH_i(s)]\bigr) \dt \Bigr).
\end{equation}
Note that the regression coefficient $\tilde{\beta}$ and baseline hazard $\tilde{\lambda}_0(t)$ in the approximation~\eqref{eq:centralappr} differ from the original ones in Equation~\eqref{eq:central}. Also note the conditioning on $T_i \geq t$ in~\eqref{eq:centralappr}, rather than on $T_i >s$, by definition of the hazard. We expect the approximation in Equation~\eqref{eq:centralappr} to be accurate when $\beta$ and the baseline hazard are not too large and when $X(t)$ is not too variable. The approximation in~\eqref{eq:centralappr} leads to the following proposal for what we call {\emph{landmarking 2.0}}:
\begin{itemize}
  \item Define and fit a working Gaussian process with trend $\mu(t)$ and covariance matrix $C(t_1,t_2)$ of the observed $X_{ij}$. Note that it is not assumed that $X_i(t)$ follows a Gaussian process, although a transformation of the longitudinal measurements in order to make it approximately normal before fitting the model is probably wise anyway;
  \item Use the fitted Gaussian process to estimate $E\{ X_i(t) \given T_i \geq t, \HH_i(s) \}$ for $t \geq s$ by least squares yielding the predictable time-dependent covariate $\hat{x}_i(t \given s)$ at each of the event time points in the data;
  \item Fit a landmark Cox model with a fixed effect of the time-dependent covariate $\hat{x}_i(t \given s)$, yielding estimates $\hat{\beta}$ and $\hat{\lambda}_0(t)$;
  \item Use the resulting landmark Cox model to obtain dynamic predictions for a new patient with observed $\HH^*(s)$ by
  \begin{itemize}
    \item Using the Gaussian process again to estimate $E\{X^*(t) \given T \geq t, \HH^*(s)\}$ for $t \geq s$ by least squares yielding the predictable time-dependent covariate $\hat x^*(t \given s)$. Let $\mu_1$ and $\mu_2$ denote the means of this fitted Gaussian process, evaluated at the observed measurement time points before $s$ and the event time points after $s$, respectively, and $\Sigma_{11}$, $\Sigma_{12}$ and $\Sigma_{22}$ the relevant sub-matrices of the variance-covariance matrix at the collection of those time points, then
\[
    \hat x^*(t \given s) = E\{X^*(t) \given T \geq t, \HH^*(s)\} = \mu_2 + \Sigma_{12}^\top \Sigma_{11}^{-1}
        (\HH^*(s) - \mu_1);
\]
    \item Calculating the predicted hazard increments $\hat{\lambda}_0(u) \exp\{ \hat{\beta} \hat x^*(u \given s) \}$ for each event time point $u$ between $s$ and $s + w$ in the data;
    \item The estimated conditional survival probability is given by
\[
    \hat{\pi}(s+w | s) = \exp\Bigl[ - \sum_{s < u \leq s + w} \hat{\lambda}_0(u) \exp\{\hat{\beta} \hat x^*(u \given s)\} \Bigr].
\]
  \end{itemize}
\end{itemize}

\noindent%
The approach is obviously more complex than na\"{\i}ve landmarking, but it is computationally considerably less challenging than joint modelling, because it avoids latent variables and integration over random effects. It gives a robust estimate of the survival given the predictable $\hat{x}(t \given s)$.  Note that the first two steps above could be replaced by any other approach that would give estimates of $\hat{x}_i(t \given s)$ for $t \geq s$. Later we will use revival modeling~\citep{dempsey2018survival} for this purpose. Landmarking 2.0 might be less efficient than the joint model, but it allows closer inspection and direct modelling of the trajectories of the survivors before estimating the regression parameters of the survival model. Note that the BLUP approach~\cite{sweeting2017use, paige2017use, paige2018landmark} is similar in spirit, but uses $\hat x_i(s \given s)$ rather than $\hat x_i(t \given s)$ in the landmark model.

As a working longitudinal model, we propose to take a variance components approach related to an autoregressive model, as also used in Dempsey \& McCullagh~\cite{dempsey2018survival}. This involves specifying a model for the trend $\mu(t)$ and one for the temporal covariance $C(t_1,t_2) = \cov(X(t_1), X(t_2))$. For the temporal covariance we follow~\cite{dempsey2018survival} by taking as variance components a between individuals variance $\sigma^2_1$, a within individuals variance $\sigma^2_2$ with a temporal correlation $\exp(-\lambda |t_1-t_2|)$, and a white noise error component $\sigma^2_3$, leading to
\[
    C(t_1, t_2) = \sigma_1^2 + \sigma_2^2 \exp(-\lambda |t_1-t_2|) + \sigma_3^2 {\bf{1}}\{ t_1=t_2 \}.
\]
Other options, depending on the fit of the model, are of course possible. The mode can be fitted with standard software for linear mixed effects models, such as the R package \textbf{nlme}.

\subsection*{Revival}

The working Gaussian process is not the only way to obtain estimates $\hat{x}(t \given s)$ to be used in the landmark Cox model. Another interesting way to achieve the same goal is to base $\hat{x}(t \given s)$ on the revival approach of~\cite{dempsey2018survival}. In this approach a longitudinal model is used for the biomarker, backwards in time from the time of death of the individual. Bayes' formula can be used to obtain dynamic predictions of $\CS_i(s+w \given s)$. A problem that has to be dealt with is the possibility of censoring, i.e., of not observing the time of death. We follow here the approach suggested by the commentary on the paper by Dempsey \& McCullagh by van Houwelingen~\cite{van2018commentary}. We start by giving some details on the model, then show how the model can be used to obtain dynamic predictions, and continue to illustrate how the same model can be used to obtain estimates of the predictable time-dependent covariate $\hat{x}(t \given s)$, to be used in landmarking 2.0.

\subsubsection*{The revival model}

Following~\cite{van2018commentary}, we start by defining an observation limit $\tau$, and denote the subset of ``dead`` subjects by those that died before $\tau$, and the subset of ``survivors'' by those that were alive at time $\tau$ (including those that are observed to die after time $\tau$). We define and fit separate models for the longitudinal markers of the dead subjects and the survivors. Subjects that were censored before $\tau$ are not included in either model. They are used later on when obtaining and assessing dynamic prediction probabilities. For subject $i$ belonging to the subset of dead people, let $t_i$ be the time of death, denote $u = t_i - t$ the reverse time to death of subject $i$, and define the time-reversed process of subject $i$ as $Z_i(u) = X_i(t_i-u)$. The distribution of this time-reversed process may depend on subject-specific factors like age, sex, and treatment. For all subjects belonging to the subset of survivors, we denote $u = \tau - t$ the reverse time to the observation limit, and define the time-reversed process as $Z_i(u) = X_i(\tau - u)$.

\subsubsection*{Dynamic prediction using revival}

When models have been defined for the time-reversed marker processes, backwards in time from the time of death $t_i$ of the dead subjects, and from the horizon $\tau$ for the survivors, conditional probabilities $\CS_i(s+w \given s)$ can be obtained by Bayes' rule, after having obtained an estimate of the marginal conditional survival probabilities. The latter may be obtained from a Kaplan-Meier estimate, possibly stratified by covariates, or a simple baseline Cox model. Since these all yield estimates that concentrate their probability mass on the observed event time points, let $t > \tpred$ be such an event time point. Then Bayes' rule gives
\begin{equation}\label{eq:directrevival}
  P(T=t \given T > \tpred, \overline{X}(\tpred)) =
    \frac{P(\overline{X}(\tpred) \given T=t, T > \tpred) \cdot P(T=t \given T > \tpred)}
      {\sum_{u>s} P(\overline{X}(\tpred) \given T=u, T > \tpred) \cdot P(T=u \given T > \tpred)}.
\end{equation}
Note that in the above, the sum in the denominator is over all event time points $u > s, u < \tau$, plus the pre-defined horizon $\tau$. Note also that $P(\overline{X}(\tpred) \given T=t, T > \tpred)$ can be simplified to $P(\overline{X}(\tpred) \given T=t)$. With slight abuse of notation, we denote by $P$ either a discrete probability or a (joint) density. For each event time point, $P(\overline{X}(\tpred) \given T=t)$ is the joint density of the observed marker values before the landmark time point $s$, which can be obtained from the distribution of the time-reversed marker process $Z(t-u)$. For $t=\tau$, the joint density of the observed marker values before the landmark time point $s$ can be obtained from the distribution of the process $Z(\tau-u)$.

\subsubsection*{Landmarking 2.0 using revival}
\label{subsub:landmarkrevival}

Equation~\eqref{eq:directrevival} describes how to obtain dynamic prediction probabilities of survival, given observed marker values. We shall refer to this as \emph{direct} dynamic prediction using revival. The time-reversed marker processes also imply conditional distributions of the marker at time points $t>s$, given survival until time $t$ and the observations of the marker process at time points before time $s$. Here we want to extract the conditional expectations of $X(t)$, given the observed history until time $\overline{x}(s)$, and given $T \geq t$. Using the same abuse of notation ($P$ denoting either density or probability), we have
\begin{eqnarray*}
    P(X(t) = x \given T \geq t, \overline{x}(s)) &=& \int_t^\tau P(X(t) = x, T=u \given T \geq t, \overline{x}(s)) \\
        &=& \int_t^\tau P(X(t) = x \given T=u, T \geq t, \overline{x}(s)) \cdot P(T=u \given T \geq t, \overline{x}(s)) \\
        &=& \int_t^\tau P(X(t) = x \given T=u, \overline{x}(s)) \cdot P(T=u \given T \geq t, \overline{x}(s)). \\
\end{eqnarray*}
When working with Cox models or non-parametric models for the time-to-event distribution, the integral is in fact a sum over event time points $u$, including the separate time point $\tau$, representing the survivors. This implies that the conditional expectation $\hat{x}(t \given s)$ is given by
\begin{eqnarray*}
    E\left( X(t) \given T \geq t, \overline{x}(s) \right) &=&
        \int x P(X(t) = x \given T \geq t, \overline{x}(s)) \, dx \\
    &=& \sum_{t \leq u \leq \tau} E\left( X(t) \given T = u, \overline{x}(s) \right) \cdot P(T=u \given T \geq t, \overline{x}(\tpred)).
\end{eqnarray*}
The last term can be written as
\[
    P(T=u \given T \geq t, \overline{x}(\tpred)) =
    \frac{P(\overline{x}(\tpred) \given T=u) \cdot P(T=u \given T \geq t)}
      {\sum_{u^\prime \geq t} P(\overline{x}(\tpred) \given T=u^\prime) \cdot P(T=u^\prime \given T \geq t)},
\]
similar to~\eqref{eq:directrevival}, the direct revival dynamic prediction probability of dying at time $u$. Both $u$ and $u^\prime$ include $\tau$. Furthermore, the first term, $E\left( X(t) \given T = u, \overline{x}(s) \right)$ can be obtained from the joint distribution of $(\overline{X}(s), \overline{X}(s, u])$, given $T = u$. Here $\overline{X}(s, u]$ refers to the vector of $X(t)$'s for the event times $t$ in $(s, u]$. If we denote this distribution as multivariate normal with mean vector $\left(
      \begin{array}{c}
        \mu_s \\
        \mu_{su} \\
      \end{array}
    \right)$, and covariance matrix $\left(
      \begin{array}{cc}
        \Sigma_{ss} & \Sigma_{s,su} \\
        \Sigma_{su,s} & \Sigma_{su,su} \\
      \end{array}
    \right)$, then we obtain
\[
    E\left( \overline{X}(s, u] \given T = u, \overline{x}(s) \right) = \mu_{su} + \Sigma_{su,s} \Sigma_{ss}^{-1} \left( \overline{x}(s) - \mu_s \right).
\]
All this implies the following procedure to calculate $\hat{x}(t \given s) = E\left( X(t) \given T \geq t, \overline{x}(s) \right)$ based on revival, for use in landmarking 2.0: first calculate the direct dynamic prediction probabilities $P(T=u \given T \geq t, \overline{x}(\tpred))$, then loop over event time points $u>s$, including $u=\tau$, and
\begin{itemize}
  \item Calculate conditional expectations and variances, given $T=u$, of $(\overline{X}(s), \overline{X}(s,u])$, yielding expectation vector $\left(
      \begin{array}{c}
        \mu_s \\
        \mu_{su} \\
      \end{array}
    \right)$ and covariance matrix $\left(
      \begin{array}{cc}
        \Sigma_{ss} & \Sigma_{s,su} \\
        \Sigma_{su,s} & \Sigma_{su,su} \\
      \end{array}
    \right)$;
  \item Calculate $E\left( \overline{X}(s, u] \given T = u, \overline{x}(s) \right) = \mu_{su} + \Sigma_{su,s} \Sigma_{ss}^{-1} \left( \overline{x}(s) - \mu_s \right)$;
  \item Combine elements $E\left( X(t) \given T = u, \overline{x}(s) \right)$ of these with $P(T=u \given T \geq t, \overline{x}(\tpred))$ and sum over $u \in (t, \tau]$ to obtain $E\left( X(t) \given T \geq t, \overline{x}(s) \right)$.
\end{itemize}

\section{Illustration}

\begin{figure}
\centering
  \includegraphics[width=0.85\textwidth]{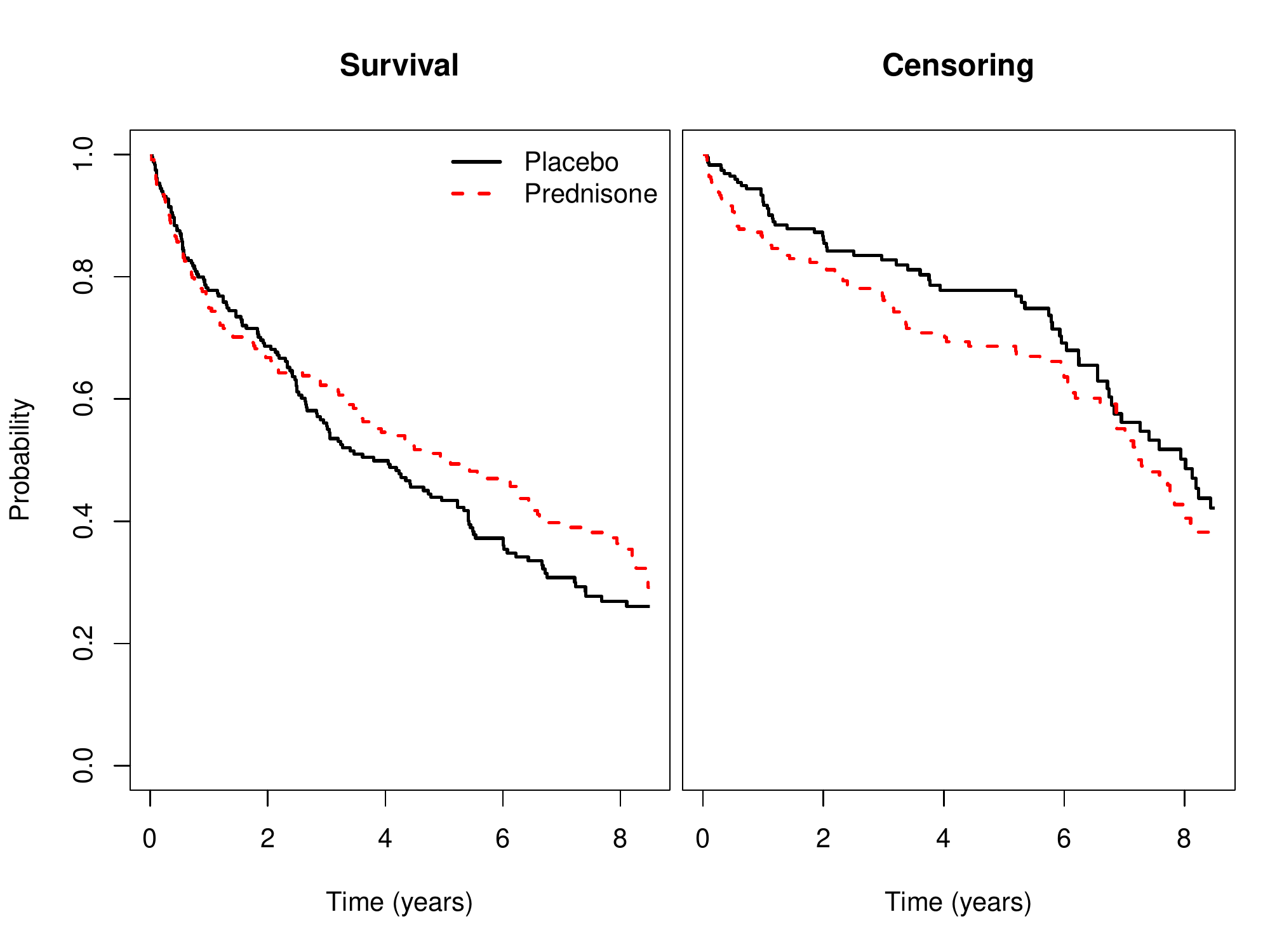}\\
  \caption{Kaplan-Meier estimates of overall survival (left) and censoring distribution (right)}\label{fig:oscens}
\end{figure}

\begin{figure}
\centering
  \includegraphics[width=0.85\textwidth]{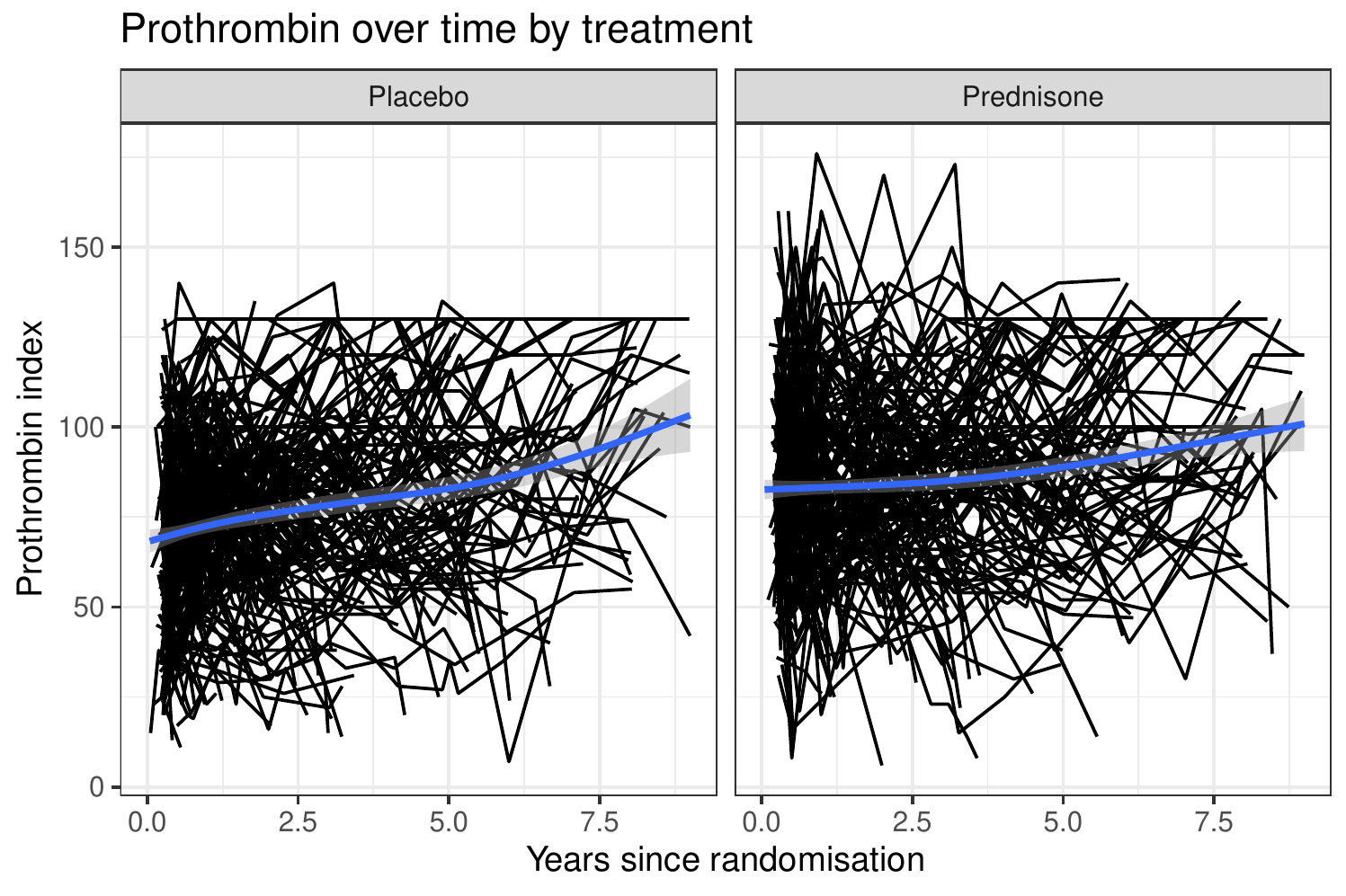}\\
  \caption{Spaghetti plot of the prothrombin values over time}\label{fig:spaghetti}
\end{figure}

We will illustrate our methods using data from the CSL-1 trial, conducted in Copenhagen in 1962-1969, randomizing patients with histologically verified liver cirrhosis to placebo or prednisone. The subset used in this paper consists of 488 patients, 251 in the prednisone and 237 in the placebo arm. Figure~\ref{fig:oscens} shows the Kaplan-Meier estimate of overall survival for all subjects in the trial, and the reverse Kaplan-Meier estimate of the censoring distribution in the trial, both by randomized treatment.
The longitudinal marker of interest is the prothrombin index, a composite blood coagulation index related to liver function, measured initially at three-month intervals and subsequently at roughly twelve-month intervals. The prothrombin measurements over time for all patients in the trial by randomized treatment are shown in Figure~\ref{fig:spaghetti}, along with a loess smoothed average.

A Gaussian process was fitted on the prothrombin measurements excluding $t=0$, where the mean $\mu(t)$ was fitted using different linear trends for the two treatment arms, but a common covariance function was used for both treatments. The estimated linear trend was $69.03 + 2.19 t$ for the placebo arm and $80.57 + 1.03 t$, with covariance parameters as shown in Table~\ref{tab:cov}.
\begin{table}
  \centering
\begin{tabular}{lll}
\hline
  Component           & Variance & Estimate\\
  				\hline
  Between individuals & $\sigma^2_1$ & 308.4 \\
  Within individuals  & $\sigma^2_2$ & 240.8 \\
  \ \ Temporal decay parameter & $\lambda$ & 0.52 \\
  White noise		  & $\sigma^2_3$ & 184.3 \\
  				\hline
\end{tabular}
  \caption{Estimated covariance parameters of the Gaussian process}\label{tab:cov}
\end{table}

\subsection*{Revival}

\begin{figure}
  \centering
  \includegraphics[width=0.85\textwidth]{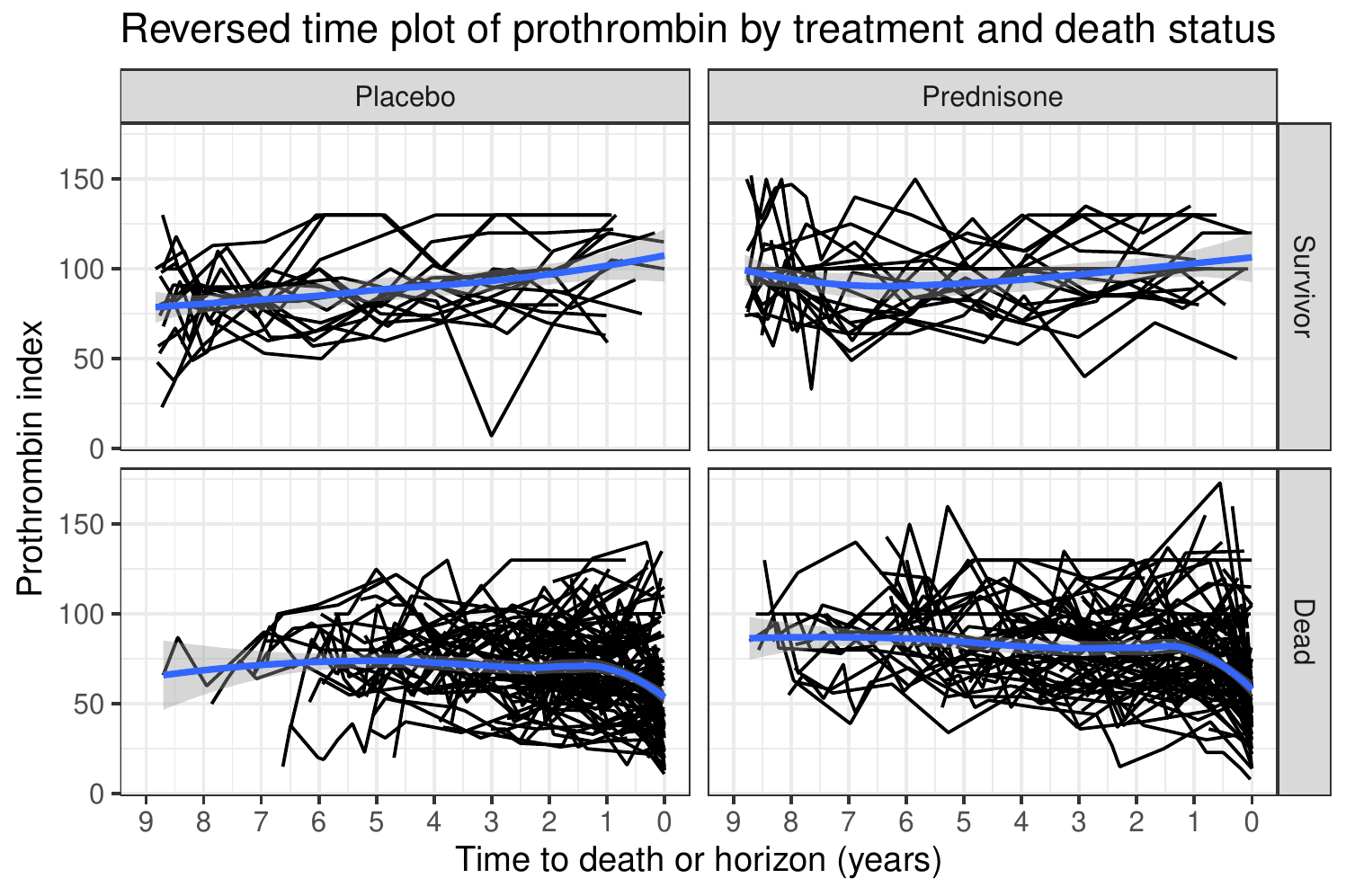}
  \caption{Spaghetti plots of the prothrombin measurements in backward time, separately for the placebo and prednisone patients, and separately for subjects that died within $\tau=9$ years (``Dead'') and subjects that were alive at nine years (``Survivor'').}\label{fig:spaghettibw}
\end{figure}

Figure~\ref{fig:spaghettibw} shows spaghetti plots of the prothrombin measurements in reverse time, separately for the placebo and prednisone patients, and separately for subjects that died within the observation limit of $\tau=9$ years and subjects that were alive at $\tau$.
Denote $A_i$ as the treatment indicator (0=placebo, 1=prednisone). A Gaussian process for the time-reversed marker process was used, with
\begin{eqnarray*}
    \E (Z_i(u) \given T=t_i, A_i) &=& \beta_0 + \beta_1 A_i + \beta_2 u + \beta_3 \log(u + \eps) + \beta_4 t_i, \\
    \cov(Z_i(u), Z_i(u^\prime) \given T = t_i) &=& \sigma_1^2 + \sigma_2^2 \exp(-\lambda |u - u^\prime|) + \sigma_3^2 {\bf{1}}\{ u = u^\prime \},
\end{eqnarray*}
for subjects that died before $\tau$ ($t_i$ being the time of death of subject $i$), and with
\begin{eqnarray*}
    \E (Z_i(u) \given T>\tau, A_i) &=& \tilde{\beta}_0 + \tilde{\beta}_1 A_i + \tilde{\beta}_2 u + \tilde{\beta}_3 \log(u + \eps), \\
    \cov(Z_i(u), Z_i(u^\prime) \given T > \tau) &=& \tilde{\sigma}_1^2 + \tilde{\sigma}_2^2 \exp(-\tilde{\lambda} |u - u^\prime|) + \tilde{\sigma}_3^2 {\bf{1}}\{ u = u^\prime \},
\end{eqnarray*}
for subjects that were alive and under follow-up at time $\tau$. For $\eps$ we took one day. The results are shown in Table~\ref{tab:reversal}.
\begin{table}
  \centering
	\begin{tabular}{llllll}
		\hline\hline
		Parameter         & Died & Censored \\
		\hline
		Intercept       & 66.39 & 95.85 \\
		$t_i$           &  1.73 & \\
		Revival $(u)$   & -1.79 & -1.39 \\
		$\ln(u+\delta)$ &  4.58 & -1.65 \\
		Prednisone      &  8.37 &  9.53 \\
        \hline
        Between individuals ($\sigma^2_1$) & 221.5 & 202.4 \\
        Within individuals ($\sigma^2_2$) & 243.6 & 191.1 \\
        \ \ Temporal decay parameter ($\lambda$) & 0.62 & 0.35 \\
        White noise ($\sigma^2_3$) & 161.9 & 161.9 \\
		\hline\hline
	\end{tabular}
  \caption{Estimates of the longitudinal time-reversed models}\label{tab:reversal}
\end{table}

\begin{figure}
	\centering
	\includegraphics[width=0.85\textwidth]{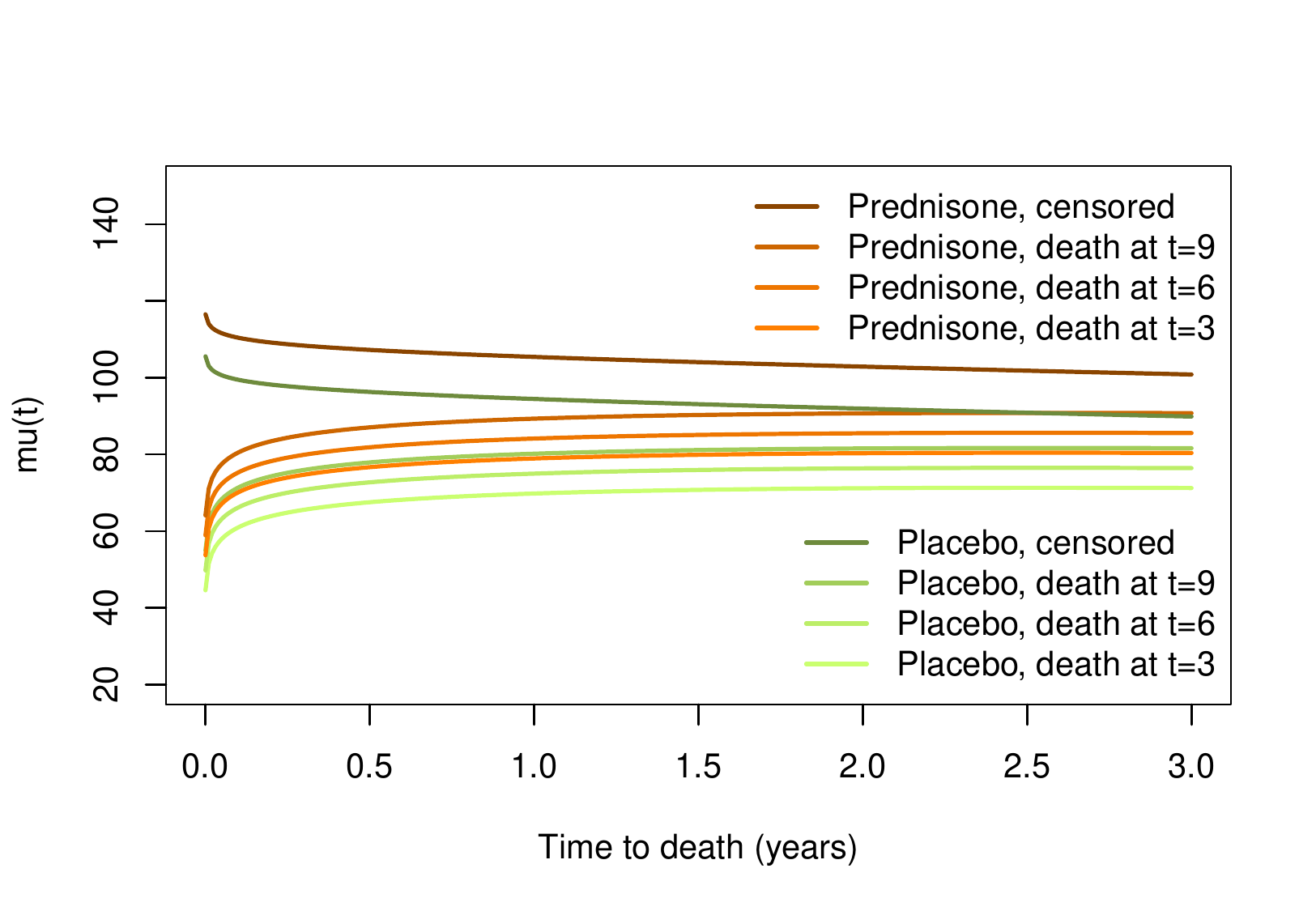}\\
  \caption{Model-based means for patients in both treatment arms, dying at 3, 6, and 9 years, and surviving until $\tau$}\label{fig:reversal}
\end{figure}

Figure~\ref{fig:reversal} illustrates the mean model, for patients in both treatment arms, dying at 3, 6, and 9 years, and surviving until $\tau$.

\subsection*{Dynamic prediction}

Our aim is to illustrate our new proposed method in obtaining dynamic prediction probabilities, and to compare these dynamic prediction probabilities with those obtained by other methods. For this purpose we fix the prediction time point at $s=3$ years and the prediction window to $w=2$ years. Using the marker values up to $s$, the following methods are considered for estimating $\CS_i(s+w \given s)$.
\begin{itemize}
  \item Joint model (JM): Joint model, where the linear mixed effects model used fixed and random intercepts (unstructured), separately for the two treatment arms, and a proportional hazards model with treatment for the survival part and piecewise constant baseline hazard, using the JM package~\citep{JM};
  \item Revival: Direct revival, using Equation~\eqref{eq:directrevival};
  \item LOCF: last observation carried forward, this is the na\"{\i}ve landmark method, where at time $s$ the last observed marker value before time $s$ is used in a Cox model;
  \item $\hat{x}(s \given s)$ (Xhats): this is the BLUP method~\cite{sweeting2017use, paige2017use, paige2018landmark}, based on the fitted working Gaussian process;
  \item $\hat{x}(t \given s)$ (Xhat): the newly proposed landmark method, with $\hat{x}(t \given s)$ based on the fitted working Gaussian process;
  \item $\hat{x}(t \given s)$ based on revival (Xhatrevival): the newly proposed landmark method, with $\hat{x}(t \given s)$ based on revival model;
\end{itemize}

Dynamic prediction probabilities were obtained by leave-one-out cross-validation; for each subject, the above models were fitted on data with the subject left out and subsequently used to obtain the predicted probability for that subject. 

\begin{figure}
  \centering
    \includegraphics[width=0.85\textwidth]{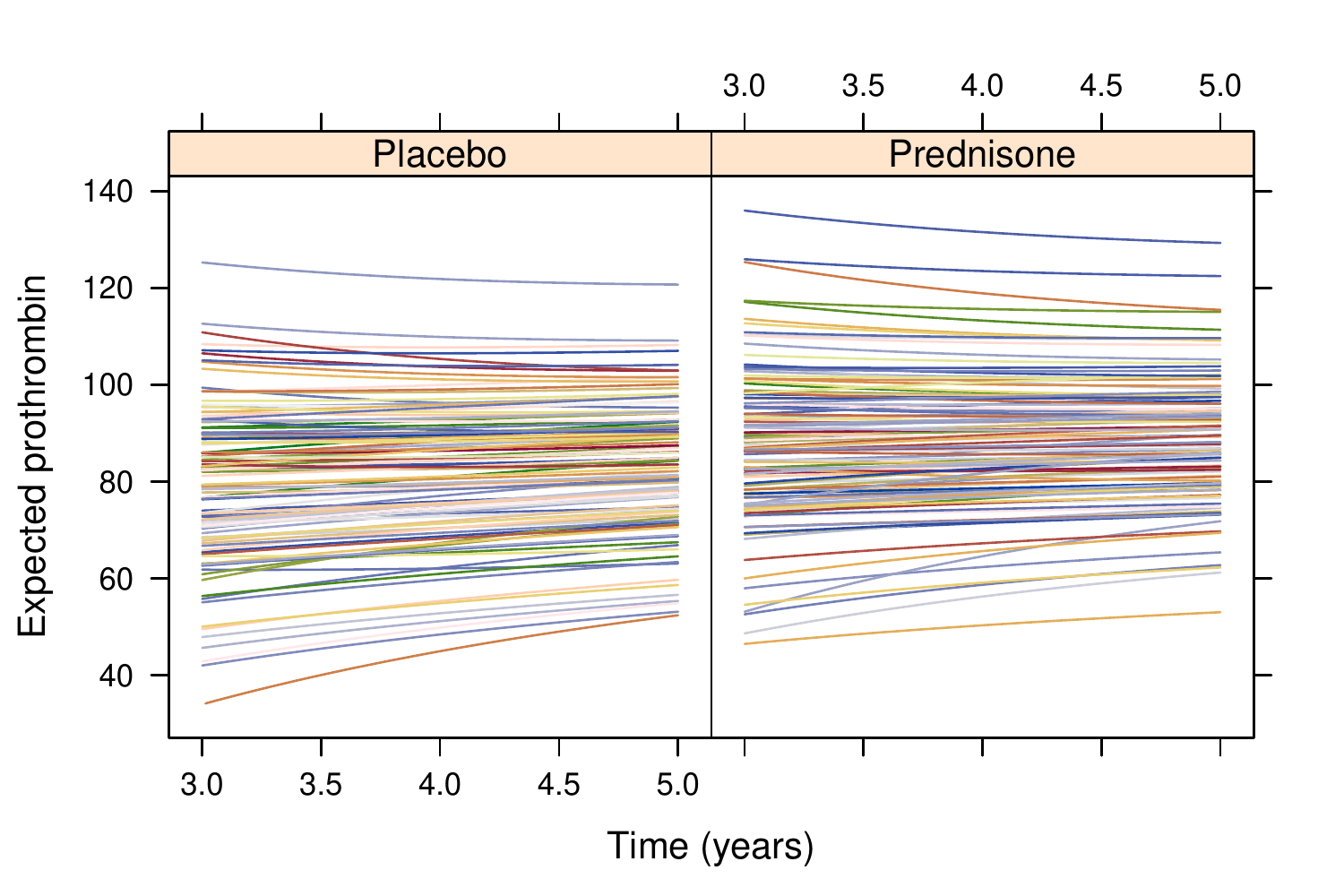} \\
    \includegraphics[width=0.85\textwidth]{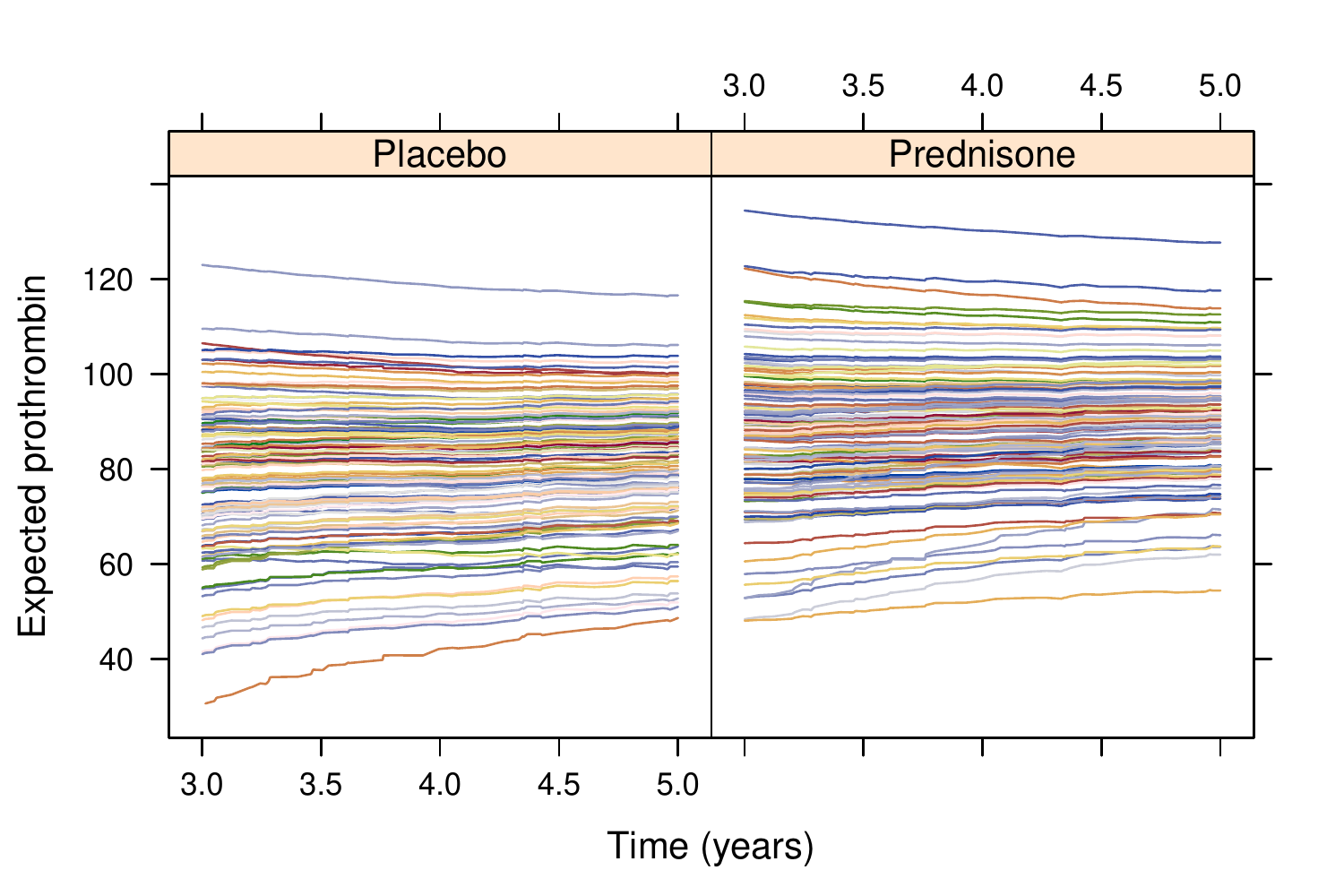}
  \caption{Evolution of $\hat{x}(t \given s)$ for $s < t \leq s+w$ for all patients by treatment, based on the Gaussian process (top row) and the revival model (bottom row)}
  \label{fig:Xhats}
\end{figure}

Figure~\ref{fig:Xhats} show the evolution of $\hat{x}(t \given s)$ for each of the patients in the CSL-1 trial, based on their observed marker values until the landmark time $s=3$ years, by randomized treatment. The $\hat{x}(t \given s)$ in the top row of Figure~\ref{fig:Xhats} are based on the fitted working Gaussian process, while those in the bottom row are based on the revival model.

\begin{figure}
  \centering
  \includegraphics[width=0.8\textwidth]{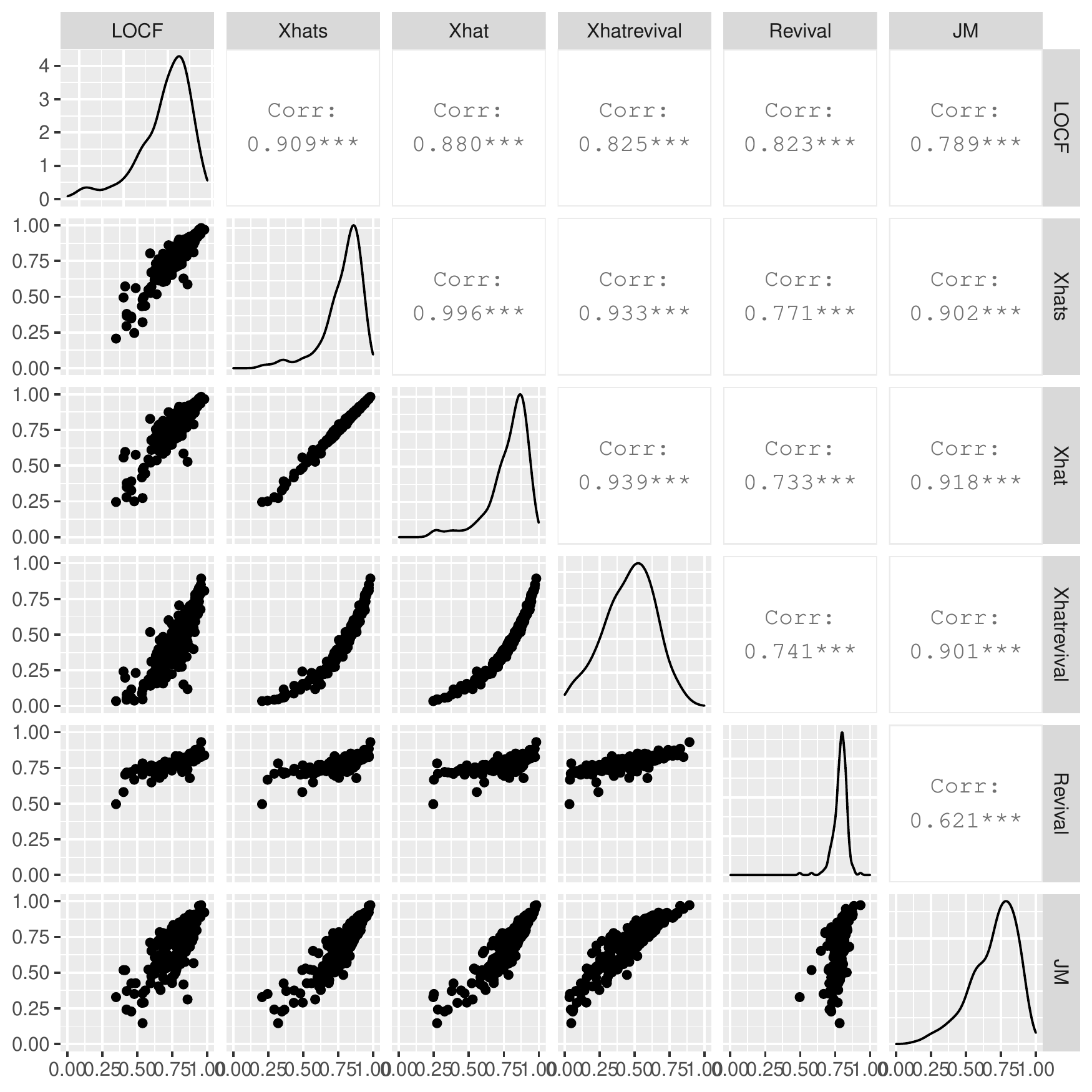}
  \caption{Matrix plot of the dynamic predictions obtained from the different approaches}
  \label{fig:comparison}
\end{figure}

Figure~\ref{fig:comparison} shows a matrix plot of the cross-validated dynamic predictions obtained from the different approaches.
Figure~\ref{fig:comparison} reveals an aspect already noted in~\cite{van2018commentary}, namely that the revival models are not well calibrated. The cross-validated prediction probabilities of the direct revival model are much more narrowly distributed around its mean than the other methods, while those of the landmarking based on revival (Xhatrevival) seem to have a somewhat lower average than the other methods. The miscalibration of the direct revival model could be due to a misspecification of the revival models (the longitudinal models in reverse time); Bayes' rule, as used in Equation~\eqref{eq:directrevival}.

In order to compare the predictive information of the different methods, we transformed the original cross-validated predicted probabilities using the complementary log-log transformation. We then entered each of the transformed cross-validated dynamic prediction probabilities in a univariate proportional hazards model in the landmark data, using administrative censoring at the horizon. The results are shown in Table~\ref{tab:comparison}.
\begin{table}
  \centering
  \begin{tabular}{lccccc}
\hline\hline
  Model                         & Beta    & SE     & $\chi^2$ & LRT \\
                                &         &        &          &  \\
\hline
  Joint model                   & 0.908 & 0.221 & 17.29 & 0.56 \\
  Direct revival                & 2.717 & 0.622 & 17.69 & 1.76 \\
  Last observation              & 0.858 & 0.238 & 12.99 & 0.17 \\
  $\hat{x}(s \given s)$         & 0.886 & 0.198 & 19.30 & 0.10 \\
  $\hat{x}(t \given s)$         & 0.874 & 0.194 & 19.45 & 0.02 \\
  $\hat{x}(t \given s)$ revival & 1.168 & 0.256 & 20.30 & --- \\
  \hline\hline
\end{tabular}
  \caption{Estimated regression coefficients, standard errors, and chi-squared statistics for the univariate ($\chi^2$) and bivariate vs univariate model with $\hat{x}(t \given s)$ revival (LRT) }\label{tab:comparison}
\end{table}
It can be seen that landmarking 2.0 with $\hat{x}(t \given s)$ based on revival as predictable time-dependent covariate has the highest univariate $\chi^2$ value ($\chi^2$ column). Landmarking 2.0 ($\hat{x}(t \given s)$) and the BLUP method ($\hat{x}(s \given s)$) are very close with respect to their univariate $\chi^2$ value. The fact that direct revival is not well calibrated is also evident from this table, with an estimated regression coefficient of 2.72. In the other methods the calibration slope is acceptable, but it must be noted that calibration in the large was not assessed here. For that, a parametric model like Poisson or Weibull~\citep{van2000validation, van2011dynamic, crowson2016assessing} could be used. After having selected landmarking 2.0 with $\hat{x}(t \given s)$, we fitted bivariate proportional hazards models with the cloglog-transformed landmarking 2.0 with $\hat{x}(t \given s)$ cross-validated dynamic prediction probabilities along with each of the other transformed cross-validated dynamic prediction probabilities. The column LRT reports the likelihood ratio test statistic of each of the bivariate Cox models, compared with the univariate Cox model with only $\hat{x}(t \given s)$ revival. The direct revival dynamic prediction probabilities gives the highest LRT, but its value is not dramatic; adding it would not yield statistical significance at the 5\% level.

Finally Table~\ref{tab:allLMs} reports the cross-validated prediction errors (both Brier and Kullback-Leibler, KL) and the percentage of prediction error reduction with respect to the null model, containing no covariates. For the revival models, the calibrated dynamic prediction probabilities (the model-based prediction probabilities based on the univariate Cox models described above) were used.
\begin{table}
  \centering
  \begin{tabular}{lcc}
\hline\hline
  Model                         & \multicolumn{2}{c}{Prediction error} \\
                                & Brier          & KL \\
\hline
  Null model                    & 0.1683         & 0.5206 \\
  Joint model                   & 0.1649 (2.1\%) & 0.5048 (3.0\%) \\
  Direct revival                & 0.1565 (7.0\%) & 0.4858 (6.7\%) \\
  Last observation              & 0.1585 (5.8\%) & 0.4932 (5.3\%) \\
  $\hat{x}(s \given s)$         & 0.1549 (8.0\%) & 0.4797 (7.9\%) \\
  $\hat{x}(t \given s)$         & 0.1549 (8.0\%) & 0.4791 (8.0\%) \\
  $\hat{x}(t \given s)$ revival & 0.1536 (8.7\%) & 0.4751 (8.7\%) \\
  \hline\hline
\end{tabular}
  \caption{Cross-validated Brier and Kullback-Leibler (KL) prediction errors of different prediction methods; in brackets after the prediction errors are the percentage reduction of prediction error, compared to the null model}\label{tab:allLMs}
\end{table}

\section{Discussion}

The landmarking principle that ``prediction should depend only on the past and nothing but the past in a transparent way'' firmly stands. Nevertheless there is a lot to learn from the ``future of the past''. We have incorporated this in landmarking~2.0 by defining a predictable time-dependent covariate to be used in a time-dependent Cox model, from the landmark time point onwards until the prediction horizon. This predictable time-dependent covariate at time $t$ is defined as the conditional expectation of the biomarker, given alive at time $t$ and given the observed biomarker values before the landmark time point, and could be determined on the basis of an underlying Gaussian process or on a reverse-time model (``revival''). The proposed procedure is more computer intensive than landmarking~1.0, but still considerably less so than joint models, because integration over random effects is avoided. In the application we considered we found that landmarking 2.0, especially when combined with revival, showed the best predictive performance, but it should be emphasized that this was just one application. We do not claim that landmarking 2.0 with revival is always the best performing procedure; more study and experience is needed to better understand the relative advantages and disadvantages of different approaches.

The good performance of revival is interesting, but it should be used with caution. First, revival methods seems to work best when the biomarker shows a marked increase or decrease in value towards the event time point. Figure~\ref{fig:spaghettibw} and the model-based version, Figure~\ref{fig:reversal}, show that this marked decrease of prothrombin values takes place only two months before death, which means that this decrease is hard to foresee after more than two months, and therefore its use in long-term prediction (also two years as used in our application) may be more limited than it seems on first instance. The revival models also need calibration. This point was already raised in~\cite{van2018commentary}; through Bayes' rule in Equation~\eqref{eq:directrevival} any misspecification in the longitudinal revival models translate in possible miscalibration of the conditional event probabilities. The effect of the misspecification of the longitudinal revival models in landmarking 2.0 combined with revival is more obscured and hard to judge. We advise to always calibrate both direct and indirect revival models in practice.

Extension to higher-dimensional biomarkers is possible in principle, by simultaneously fitting Gaussian processes to the biomarkers. This requires more thinking how to handle the correlation between the biomarker components. The extension to higher-dimensional biomarkers is easier for landmarking 2.0 than for joint models, because for the latter approach dealing with the (typically higher-dimensional) random effects becomes comparatively much more difficult.

In this paper we covered the situation of dynamic prediction based on a biomarker, repeatedly measured over time. Analysis and prediction with this type of time-dependent covariates is often performed using joint models. Another common situation of dynamic prediction with time-dependent covariates concerns the case where the time-dependent covariate is a binary covariate, most commonly changing from 0 to 1 over the course of time. Examples include prediction of survival based on the occurrence of some intermediate event like relapse, progression or response to treatment. In that context the time-dependent covariate always starts as $X(t)=0$ at $t=0$ and cannot revert from the value 1 back to 0. Then dynamic prediction at prediction time $s$ is of most interest when $X(s)=0$ (which implies $\HH(s) \equiv 0$). One approach, the equivalent of the joint model approach for longitudinally measured biomarkers, is multi-state models, in the present case with states 0 (alive and $X(t)=0$), 1 (alive and $X(t)=1$), and 2 (dead). In that case, Equation~\eqref{eq:centralappr} can also be used, and $\E[X(t) \given T \geq t, \HH(s) \equiv 0]$ is given by the prevalence probability $\pi_{01}(s, t) = \frac{P_{01}(s, t)}{P_{00}(s, t) + P_{01}(s, t)}$, where $P_{gh}(s, t)$ are the \emph{transition probabilities} in the multi-state model, see Section 2.1 of~\cite{putter2017understanding}. The conditional survival probability can also be expressed directly as one minus the transition probability $P_{02}(s, t)$, which can also be calculated in the multi-state model. The added value of landmarking 2.0 could be that the result does not depend on the Markov assumption, see also~\cite{van2008dynamic}. It would be of interest to study this case further.

\subsection*{Acknowledgement}

Michael Sweeting is gratefully acknowledged for help in fitting Gaussian processes in R.

\subsection*{Data and code availability}

The code used to perform the analyses in this paper is available on \url{https://github.com/survival-lumc/Landmarking2.0}. Data is publicly available in the  \textbf{joineR} package~\cite{joineR}.

\bibliographystyle{WileyNJD-AMA}
\bibliography{lmjoint}

\end{document}